\documentclass[%
 reprint,longbibliography,preprintnumbers,
nofootinbib,
 amsmath,amssymb,
 aps
]{revtex4-1}
\pdfoutput=1
\usepackage{graphicx}
\usepackage[utf8]{inputenc}
\usepackage{flushend}
\usepackage{dcolumn}
\usepackage{bm}
\usepackage{balance}

\usepackage[normalem]{ulem}

\usepackage[colorlinks = true,
            linkcolor = blue,
            urlcolor  = blue,
            citecolor =green,
            anchorcolor = blue]{hyperref}
\usepackage{verbatim}
\usepackage{color,ulem}
\usepackage[english]{babel}

\usepackage[utf8]{inputenc}
\input Starburst.fd
\newcommand*\initfamily{\usefont{U}{Starburst}{xl}{n}}\initfamily

\newcommand{\beq}{\begin{eqnarray}}
\newcommand{\eeq}{\end{eqnarray}}
\usepackage{amsmath}
\usepackage{tikz}
\usetikzlibrary{decorations.pathmorphing}
\usetikzlibrary{shapes.misc}
\tikzset{cross/.style={cross out, draw=black, minimum size=8*(#1-\pgflinewidth), inner sep=0pt, outer sep=0pt},
cross/.default={1pt}}
\usetikzlibrary{patterns,math}
\begin{document}

\title{\Large  Reply to Shvaika et al.: Presence of a boson peak in anharmonic phonon models with Akhiezer-type damping}

\author{\textbf{Matteo Baggioli}$^{1,2}$}%
 \email{b.matteo@sjtu.edu.cn}
\author{\textbf{Alessio Zaccone}$^{3,4}$}%
 \email{alessio.zaccone@unimi.it}
 
 \vspace{1cm}
 
\affiliation{$^{1}$Wilczek Quantum Center, School of Physics and Astronomy, Shanghai Jiao Tong University, Shanghai 200240, China}
\affiliation{$^{2}$Shanghai Research Center for Quantum Sciences, Shanghai 201315.}
\affiliation{$^{3}$Department of Physics ``A. Pontremoli'', University of Milan, via Celoria 16,
20133 Milan, Italy.}
\affiliation{$^{4}$Cavendish Laboratory, University of Cambridge, JJ Thomson
Avenue, CB30HE Cambridge, U.K.}

\begin{abstract}
We reply to the Comment by Svhaika, Ruocco, Schirmacher and collaborators. There were two accidental mistakes in our original paper~\cite{PhysRevLett.122.145501}, which have been now corrected. All the physical conclusions and results of the original paper, including the prediction of boson peak due to anharmonicity, remain valid in the corrected version.
\end{abstract}

\maketitle
In Ref.\cite{PhysRevLett.122.145501}, we computed the vibrational density of states (VDOS) for acoustic phonon vibrations:
\begin{equation}
    \omega^2(q)\,=\,\Omega(q)^2-i \,\omega\, \Gamma(q)
\end{equation}
with diffusive quadratic damping $\Gamma(q) \sim q^2$, as derived by Akhiezer in 1939 based on anharmonicity for crystals~\cite{Akhiezer} (for an accessible derivation see pp. \textit{366-367} in~Ref.\cite{Chaikin}), and proposed by Allen and Feldman for disordered systems based on numerical calculations~\cite{Allen_1999}. As pointed out in the Comment by Svhaika, Ruocco, Schirmacher and collaborators \cite{shvaika2021absence}, our Eq.(7) in \cite{PhysRevLett.122.145501} contains an obvious mathematical inconsistency due to the absence of the 3D volume factor $q^2 dq$. Moreover, in \cite{PhysRevLett.122.145501}, we assumed the low-energy phonon dispersion relation $\Omega(q)=v q$ which does not take into account the existence of a BZ in the solid (or pseudo-BZ, for glasses) or at least the ubiquitous bending of the acoustic phonon branch towards larger wave-vector. In this Reply, we correct these issues and show that the main physical conclusions of \cite{PhysRevLett.122.145501} remain unchanged\footnote{The corrected model appeared already in~\cite{PRR}.}.
Neglecting the longitudinal modes, the correct expression for the vibrational density of states (VDOS) is given by:
\begin{equation}
    g(\omega)\,=\,-\,\frac{2\,\omega}{\pi\,q_D^3}\,\int_0^{q_D}\,\mathrm{Im}\, \mathrm{G}(\omega,q)\,q^2\,dq \label{dd}
\end{equation}
where:
\begin{equation}
    \mathrm{G}(\omega,q)\,=\,\frac{1}{\omega^2\,-\,v^2\,q^2\,+\,a_4\,q^4\,+\,i\,\omega\,D\,q^2}\label{bb}
\end{equation}
is the propagator given by Effective Field Theory (EFT) as a hydrodynamic expansion in powers of $q$ (where odd powers of $q$ are forbidden by symmetry).
This propagator is identical to the one of ~\cite{PhysRevLett.122.145501} 
with the only difference of the higher-order (quartic) term in $q$, with the phenomenological coefficient $a_4$,
which was absent in \cite{PhysRevLett.122.145501}.
This term allows for the correct description of the  bending of the acoustic branch upon approaching the BZ boundary. As in any effective field theory, Eq.\eqref{bb} is derived in an expansion in $q$ up to $4^{th}$ order which is enough to parametrize the first correction related to the bending of the acoustic branches. Eq.\eqref{bb} cannot be extrapolated to very large wave-vectors $q\gg 1$ (and in particular beyond the point at which the dispersion relation flattens) since higher order corrections would definitely enter therein, but it is a good approximation for small enough wave-vector where the BP appears.\\

\begin{figure}
    \centering
    \includegraphics[width=\linewidth]{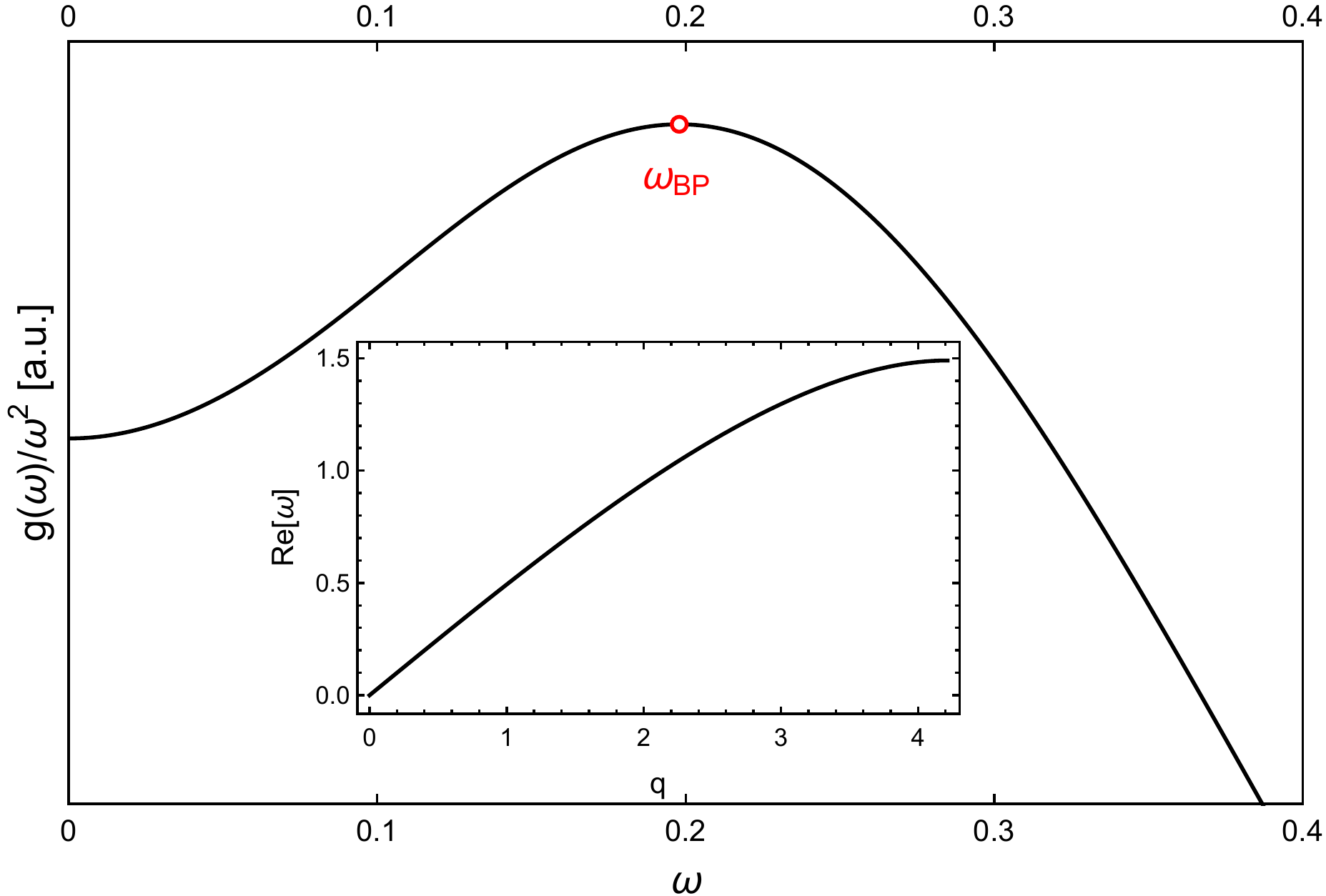}
    \caption{A concrete demonstration of the presence of a BP anomaly $\omega_{BP}$ arising from Eqs.\eqref{dd}-\eqref{bb}. Here $v=0.5,a_4=0.007,D=0.012,q_D=5.95$. Finally, $\omega_{BZ}\approx 1.5 \gg \omega_{BP}$. The inset shows the acoustic dispersion relation used to compute the VDOS as given by Eq.\eqref{bb}, where the quartic term is responsible for the flattening of the dispersion relation upon approaching the BZ boundary.}
    \label{fig:1}
\end{figure}
\begin{figure*}[ht]
    \centering
    \includegraphics[width=0.33\linewidth]{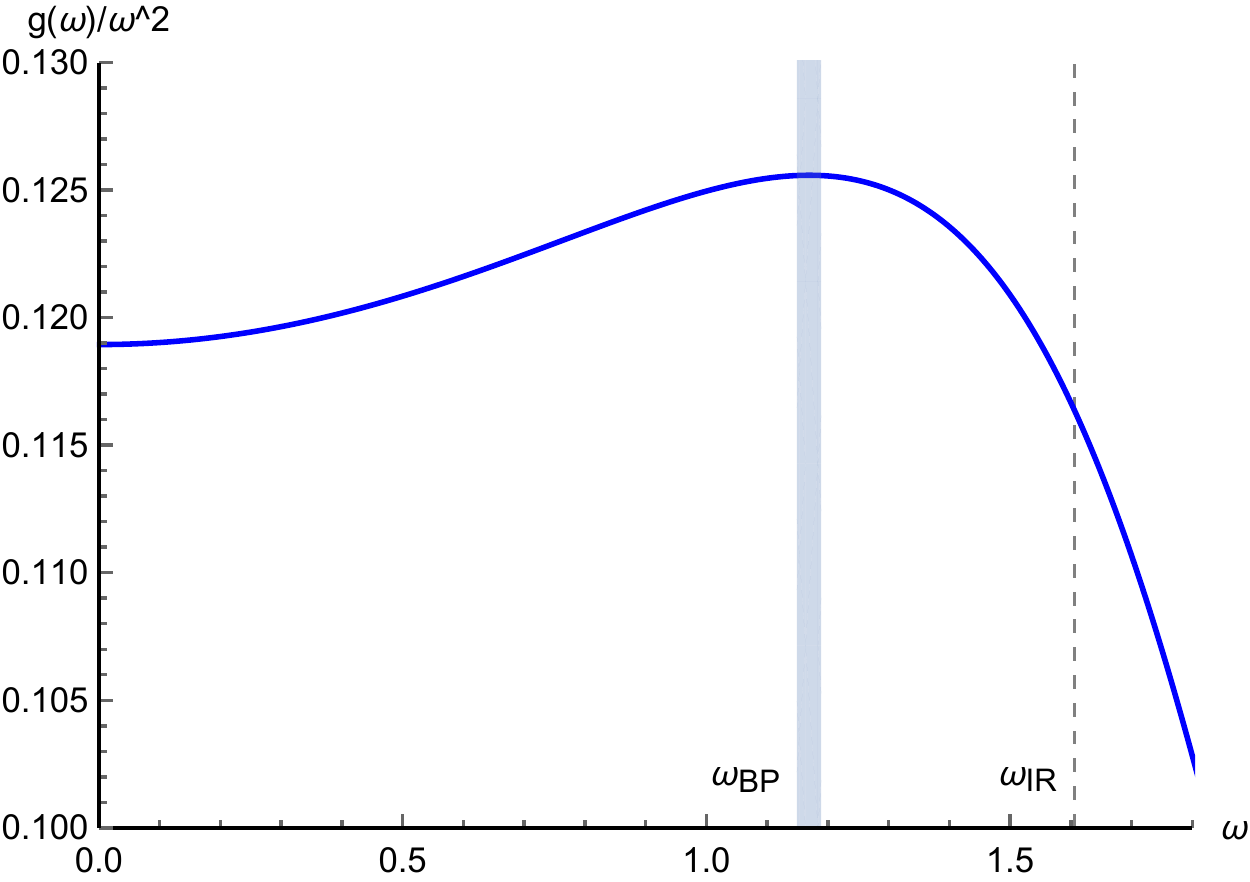}\quad 
    \includegraphics[width=0.33\linewidth]{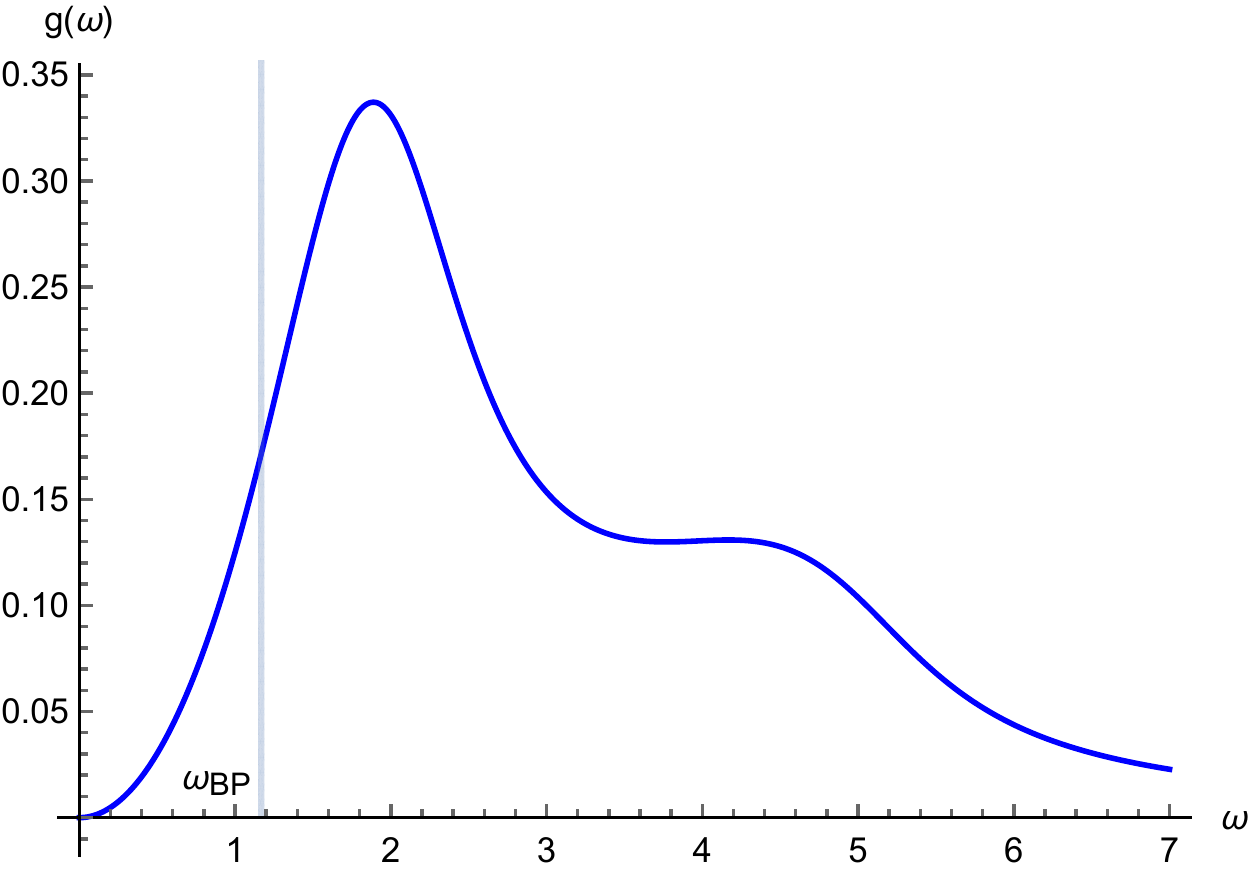}\quad
    \includegraphics[width=0.24\linewidth]{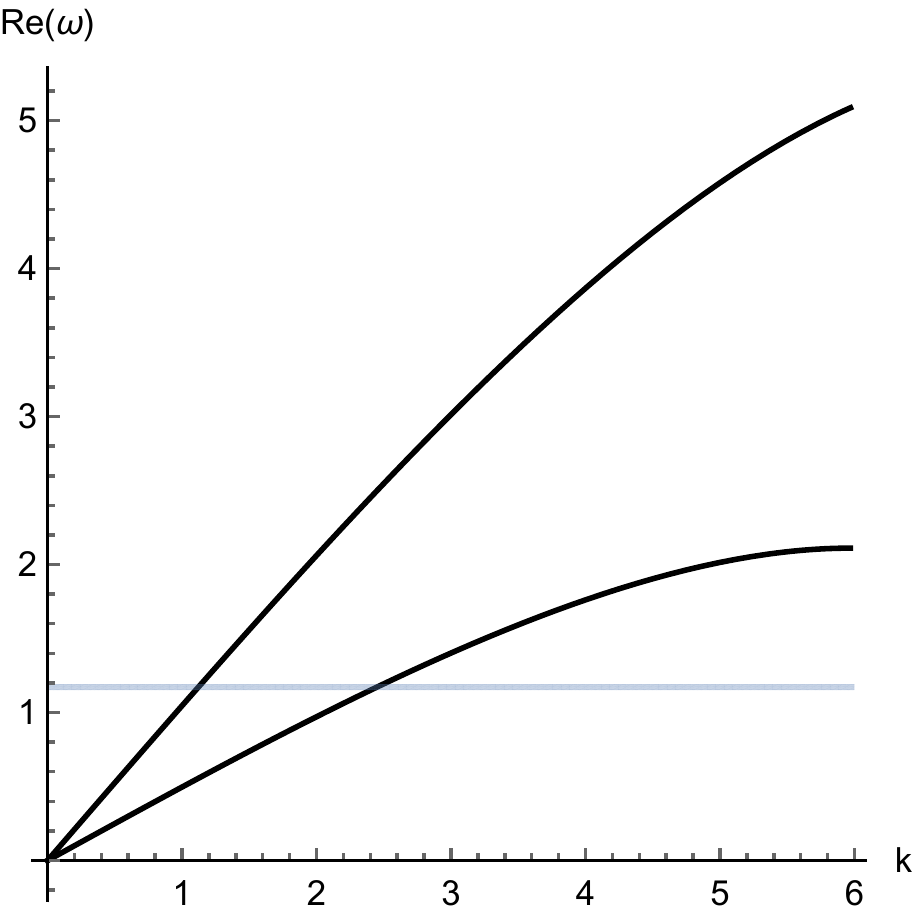}
    \caption{\textbf{Left: }the Debye-normalized density of states. \textbf{Center: }the vibrational density of states (VDOS). \textbf{Right: }the dispersion relation of transverse  (lower curve) and longitudinal (upper curve) acoustic phonons. The gray lines indicate the position of the BP. In these plots  $v_T=0.5,a_4^T=0.003,D_T=0.045,v_L=1.051,a_4^L=0.01,D_L=0.05,k_D=5.95$. Using these parameters, we have $\omega_{BP}=1.17,\omega_{IR}^T=1.6,\omega_{BZ}^T=2.11$ in the same arbitrary units.}
    \label{fig:panel}
\end{figure*}
Back to our topic, Eq~\eqref{dd} combined with Eq.\eqref{bb} predicts a boson peak anomaly in the Debye normalized density of states $g(\omega)/\omega^2$, which is well separated from the edge of the BZ, $\omega_{BZ}$, as shown in Fig.\ref{fig:1}. The emergence of the boson peak is due to the anharmonic Akhiezer-type phonon damping $\Gamma = D q^{2}$ in the propagator Eq.\eqref{bb}, without which no boson peak can be predicted. This is the same finding and main conclusion of Ref.~\cite{PhysRevLett.122.145501}. We emphasize that the quartic term in Eq.\eqref{bb} itself would not lead to the presence of any BP in the reduced density of states but only of a well-defined van Hove singularity corresponding to the flattening of the acoustic branch $d \omega/dq=0$. As evident from Fig.\ref{fig:1}, once the diffusive damping term is introduced in Eq.\eqref{bb}, a BP anomaly appears in the VDOS at frequencies much smaller than the BZ boundary (in the concrete example, $\omega_{BP}\sim 0.2$ and $\omega_{BZ}\sim 1.5$), proving the fundamental role of the Akhiezer damping term.\\

Another important result of Ref.~\cite{PhysRevLett.122.145501} was the prediction that the boson peak frequency $\omega_{BP}$ is close to the frequency of the Ioffe-Regel crossover $\omega_{IR}$ (at which acoustic phonon propagation turns from ballistic into diffusive-like) in glasses\footnote{Notice that this correlation has never been observed so far in systems lacking structural disorder but presenting a well-defined boson peak. Interestingly, both these scenarios, BP frequency correlated with IR frequency (Fig.\ref{fig:1}) and not (Fig.\ref{fig:panel}), are captured by our toy model.}, as ubiquitously observed in simulations~\cite{Tanaka} and experiments~\cite{Ruffle}.
Also this prediction of $\omega_{BP} = \mathcal{O} (\omega_{IR})$ remains confirmed in the corrected version of the model, as shown by sample calculations in
Fig.~\ref{fig:panel}. There, in order to make the model slightly more realistic we have also included the dispersion relation of the longitudinal branch. In the concrete example, it is found indeed that $\omega_{IR}/\omega_{BP} = 1.3$.\\

Hence, there were two accidental omissions in Ref.~\cite{PhysRevLett.122.145501}: (i) the metric factor $q^2$ in the sum/integration over momentum $q$ (cfr. Eq. (6) of \cite{PhysRevLett.122.145501} with Eq.~\eqref{dd} above) and (ii) the higher-momentum $q^{4}$ correction to the acoustic phonon dispersion relation (cfr. denominator of Eq. (3) of \cite{PhysRevLett.122.145501} with Eq.~\eqref{bb} above).\\

Nevertheless, as proven in this Reply, the main physics and all the conclusions of Ref.~\cite{PhysRevLett.122.145501} continue to hold and remain unaffected. Namely, the occurrence of a BP anomaly induced by a diffusive Akhiezer damping term is still a valid and important theoretical explanation, especially for ordered systems where other boson peak models based on disorder cannot be applied, as well as for amorphous solids at finite $T$ where athermal ``harmonic disorder'' theories fail to capture the important contribution of anharmonicity. 

In Summary, in this Reply, we demonstrate the \textit{presence of a boson peak in anharmonic phonon models with Akhiezer-type damping}, which invalidates all the physical conclusions and arguments presented in the Comment by Svhaika et al.~\cite{shvaika2021absence} and supports the validity of our original idea~\cite{PhysRevLett.122.145501}. The corrected Mathematica code can be downloaded from the ancillary file at Ref.~\cite{Reply}.
\subsection*{Acknowledgments} 
We thank the authors of ~\cite{shvaika2021absence} for pointing out the mathematical mistakes in~\cite{PhysRevLett.122.145501} and for motivating us to revisit our analysis.
A.Z. acknowledges financial support from US Army Research Office, contract nr. W911NF-19-2-0055. M.B. acknowledges the support of the Shanghai Municipal Science and Technology Major Project (Grant No.2019SHZDZX01).
\bibliographystyle{apsrev4-1}

\bibliography{refs}

\end{document}